\documentstyle[aps, array, manuscript]{revtex}
\tightenlines
\draft
\begin{document}
\title{Scalar Metric fluctuations in Space Time Matter inflation}
\author{
$^{1,2}$ Mariano Anabitarte\footnote{
E-mail address: anabitar@mdp.edu.ar}
and $^{1,2}$Mauricio Bellini\footnote{
E-mail address: mbellini@mdp.edu.ar}}
\address{
$^1$Departamento de F\'{\i}sica,
Facultad de Ciencias Exactas y Naturales,
Universidad Nacional de Mar del Plata,
Funes 3350, (7600) Mar del Plata, Argentina.\\
$^2$ Consejo Nacional de Ciencia y Tecnolog\'{\i}a (CONICET).}

\vskip .5cm
\maketitle
\begin{abstract}
Using the Ponce de Leon background metric, which
describes a 5D universe in an apparent vacuum: $\bar{G}_{AB}=0$,
we study the effective 4D evolution of both, the inflaton and gauge-invariant
scalar metric fluctuations, in the recently introduced model
of space time matter inflation.
\end{abstract}
\vskip .2cm
\noindent
%Keywords: cos.fth.ctg \\
Pacs numbers: 04.20.Jb, 11.10.kk, 98.80.Cq \\
\vskip 1cm
\section{Introduction and motivation}

The idea that the universe passed through an inflationary expansion
in early epochs has become an integral part
of the standard cosmological model.
Postulating a period of nearly exponential growth in the primordial
universe, inflationary cosmology solved many problems which plagued previous
models of the big bang and also provides
a mechanism for the creation of primordial density fluctuations.
During inflation, the universe is dominated by the inflaton, the scalar
field whose evolution controls the dynamics of the expansion.
The first model of inflation was proposed by A. Starobinsky in 1979\cite{Sta}.
Two years later, a model with a clear motivation was developed by
Guth\cite{Guth},
in order to solve some of the shortcomings of the big bang theory, and in
particular, to explain the extraordinary homogeneity of the observable 
universe.
The scalar field state employed in the original version
of inflation is called a false vacuum, since
the state temporarily acts as if
it were the state of lowest possible energy density.
In 1983 Linde proposed chaotic inflation\cite{lin}.
In its standard
version, the inflaton field
is related to a potential with a local minimum or a gently
plateau. In this model, cosmological perturbations
are indispensable in relating early universe scenarios\cite{Muk}.
Other standard 4D models where dissipative effects are important
during the inflationary phase, were suggested more recently\cite{6,7,8}.
The evolution of gauge-invariant meric perturbations during inflation
has been well studied\cite{per}. This allows to formulate the
problem of the amplitude for the scalar metric perturbations on the
evolution of the background Friedmann-Robertson-Walker (FRW) universe
in a coordinate-independent manner at every moment in time.

On the other hand, cosmological theories with
extra dimensions are
already known to be of great importance in
cosmology\cite{uno,dos}. During the last years there were many attempts
to construct a consistent brane world (BW)
cosmology\cite{.}. The induced-matter,
or space-time-matter (STM) theory
stands out for their closeness to the Einstein's project of considering
matter and radiation as manifestations of pure geometry\cite{eins}.
The basic extension of 4D Einstein theory and the low-energy
limit of higher-dimensional theories is the modern
incarnation of noncompact 5D Kaluza-Klein theory.
In the STM theory, the conjecture is that ordinary matter and fields
that we observe in 4D are induced geometrically
by the extra dimension\cite{we}.
In this letter we develop a consistent first-order
formalism to study the inflaton and scalar metric
fluctuations\cite{ant}
in STM inflation, which was recently introduced\cite{AB}.

\section{5D Formalism}

We consider a 5D background metric, which is 3D spatially
isotropic, homogeneous and flat. In this paper
we shall consider the background metric: $dS^2_{(b)}=
\bar{g}_{AB} dx^A dx^B$
\footnote{In our conventions,
capital Latin indices run from 0 to 4.},
introduced by Ponce de Leon\cite{uno}
\begin{equation}\label{3}
dS_{(b)}^2 =  l^2 dt^2 - \left( \frac{t}{t_0} \right) ^{2p}
l^{\frac{2p}{p-1}} dr^2 - \frac{t^2}{\left( p-1 \right)^2} dl^2,
\end{equation}
where $dr^2 = dx^2+dy^2+dz^2$ and $l$ is the fifth coordinate (which
is space-like).
The background metric (\ref{3})
represents a 5D apparent vacuum $\bar{G}_{AB}=0$,
but however it is no flat. 
The absolute value for the determinant of the background
metric tensor $\bar{g}_{AB}$ is
\begin{displaymath}
\left|^{(5)} \bar{g}\right| = \left[\frac{t^{3p+1}
l^{\frac{4p-1}{p-1}}}{(p-1) t^{3p}_0} \right]^2.
\end{displaymath}

To describe the system, we consider the action
\begin{equation}                   \label{action}
I
= - {\Large\int} d^4 x \  dl \sqrt{\frac{^{(5)} \bar{g}}{^{(5)} \bar{g}_0}}
\left[ \frac{^{(5)} \bar R}{16
\pi G} + ^{(5)} {\cal L}\right],
\end{equation}
where $^{(5)} \bar R=0$ is the 5D background Ricci scalar for the
background metric (\ref{3}) and $G=M^{-2}_p$
is the Newton's constant ($M_p =1.2 \times 10^{19} \  GeV$ is the
Planckian mass).
We shall use the unities $c=\hbar=1$,
being $c$ and $h$ the speed of light and the
Planck's constant.
The Lagrangian density in (\ref{action}) is given by
\begin{equation}\label{lag}
^{(5)} {\cal L} = \frac{1}{2} g^{AB} \varphi_{,A} \varphi_{,B},
\end{equation}
which is only kinetic because we are dealing with a 5D free scalar
field $\varphi$ in an apparent vacuum state.
Furthermore, $g^{AB} = \bar{g}^{AB} + \delta g^{AB}$ is the perturbed
contravariant metric tensor and
$\bar{g}^{AB}$ is the background
contravariant metric tensor.
In this letter we shall consider $\delta g^{AB}$
as the scalar perturbations of the metric at first order, because
we are dealing with weak gravitational fields.
The diagonal perturbed metric [with respect to
the background metric (\ref{3})], is 
\begin{equation}
dS^2 =  l^2 \left( 1+2\Phi \right)dt^2 -
\left(\frac{t}{t_0}\right)^{2p}l^{\frac{2p}{p-1}} \left( 1-2\Psi
\right) dr^2 - \frac{t^2}{\left( p-1 \right)^2} \left( 1-Q
\right)dl^2,
\end{equation}
where $\Phi(t,\vec r,l)$, $\Psi(t,\vec r,l)$ and $Q(t,\vec r,l)$
are the scalar metric fluctuations.
In
particular for $\Phi=\Psi$ and $Q=2\Psi$, we obtain the following
line element:
\begin{equation}
dS^2 =  l^2 \left( 1+2\Psi \right)dt^2 -
\left(\frac{t}{t_0}\right)^{2p}l^{\frac{2p}{p-1}} \left( 1-2\Psi
\right) dr^2 - \frac{t^2}{\left( p-1 \right)^2} \left( 1-2\Psi
\right)dl^2.
\end{equation}
The contravariant metric tensor, after make
a $\Psi$-first order approximation,
is given by
\begin{eqnarray}
g^{AB} & = & {\rm diag} \left[\left(1-2\Psi\right)/l^2,
-\left(1+2\Psi\right) l^{\frac{-2p}{(p-1)}} \left(\frac{t}{t_0}\right)^{-2p},
-\left(1+2\Psi\right) l^{\frac{-2p}{(p-1)}} \left(\frac{t}{t_0}\right)^{-2p},
\right.\nonumber \\
&-& \left.
\left(1+2\Psi\right) l^{\frac{-2p}{(p-1)}} \left(\frac{t}{t_0}\right)^{-2p},
-\left(1+2\Psi\right)  \frac{(p-1)^2}{t^2}\right], \label{mut}
\end{eqnarray}
which can be written as $g^{AB} = \bar{g}^{AB} + \delta g^{AB}$, being
$\bar{g}^{AB}$ the contravariant background metric tensor.
Furthermore, the Lagrangian $^{(5)} {\cal L} = {1\over 2} g^{AB} \varphi_{,A}
\varphi_{,B}$, with the metric fluctuations $\Psi$ included in $g^{AB}$,
now can be written as
\begin{displaymath}
^{(5)} {\cal L}\left(\varphi,\psi, \varphi_{,A},\psi_{,A}\right) =
\frac{1}{2} g^{AB} \varphi_{,A} \varphi_{,B},
\end{displaymath}
where the fields $\varphi$ and $\Psi$ play the role of coordinates and
$g^{AB}$ is given by (\ref{mut}).

The relevant components of the linearized Einstein tensor are
\begin{eqnarray}
G_{tt}  &= & \frac{1}{t^2} \left[ -3l^2 \left(p-1\right) ^2
\stackrel{\star\star}{\Psi} + 9l (p-p^2) \stackrel{\star}{\Psi} +3t
(3p+1) \dot\Psi +12 (p^2+p) \Psi \right. \nonumber \\
&-& \left. 3t^2 \left( \frac{t_0}{t} \right)
^{2p} l^{\frac{-2}{p-1}} \nabla_r^2 \Psi\right], \\
G_{rr} & = & 3\left( \frac{t}{t_0} \right) ^{2p}
\frac{l^{\frac{2p}{p-1}}}{t^2 l^2} \left[ l^2(p-1)^2
\stackrel{\star\star}{\Psi} + l(p^2-1) \stackrel{\star}{\Psi} -3t^2
\ddot\Psi -5t (2p+1) \dot\Psi - 12p^2 \Psi\right] \nonumber \\
& + & 2\nabla_r^2\Psi, \\
G_{ll} & = & \frac{1}{l^2 (p-1)^2} \left[ 3l (2p^2-3p+1)
\stackrel{\star}{\Psi} -3t^2 \ddot\Psi - 15 p t \dot\Psi + 12
p(1-2p) \Psi \right.\nonumber \\
&+ & \left.\left( \frac{t_0}{t} \right) ^{2p} l^{\frac{-2}{p-1}}
t^2 \nabla_r^2 \Psi\right], \\
G_{tl} & = & -3 \left[ \frac{1}{2} \stackrel{\dot\star}{\Psi} +
\frac{(2p-1)}{t} \stackrel{\star}{\Psi} + \frac{1}{l} \dot{\Psi}\right],
\end{eqnarray}
where the overstar denotes the derivative with respect to the fifth
coordinate $l$.
On the other hand, the energy momentum tensor can also be written as
a linear one, with background components $\bar{T}_{AB}$,
plus the first order perturbations $\delta T_{AB}$:
$T_{AB}= \bar{T}_{AB} +\delta T_{AB}=\varphi_{,A}\varphi_{,B} -
g_{AB} \varphi_{,C}\varphi^{,C}$,
where
\begin{eqnarray}
\bar{T}_{AB}  & = & \left.\varphi_{,A} \varphi_{,B} - \bar{g}_{AB}
\varphi_{,C}\varphi^{,C}\right|_{{\rm background}} =0, \\
\delta T_{AB} & = & - \delta g_{AB} \varphi_{,C} \varphi^{,C}.
\end{eqnarray}
Notice that all the terms $\delta T_{AB} $ (with $A\neq B$) are
zero, because $\delta g_{AB}$ is diagonal.
The
relevant perturbed components are
\begin{eqnarray}
\delta T_{tt} & = & 2\Psi (p-1)^2 \frac{l^2}{t^2}
(\stackrel{\star}{\varphi})^2, \label{eq10} \\
\delta T_{rr} & = & - \frac{6\Psi}{l^2}
\left(\frac{t}{t_0}\right)^{2p}l^{\frac{2p}{p-1}}(\dot\varphi)^2, \label{eq11}\\
\delta T_{ll} & = & - \frac{2\Psi t^2}{(p-1)^2 l^2} (\dot\varphi)^2, \label{eq12}
\end{eqnarray}
which are nonlinear. In order to solve the perturbed and background
Einstein equations,
we consider the following
semiclassical expansion for the inflaton field:
$\varphi(t,\vec r,l) =
\varphi_{(b)}(t,l) + \delta\varphi(t,\vec r,l)$, being
$\varphi_{(b)}$ the background inflaton field and $\delta\varphi$
the quantum fluctuations such that $\left<\delta\varphi\right>=0$.
If we take into account this semiclassical expansion in
in eqs. (\ref{eq10}), (\ref{eq11}) and (\ref{eq12}), 
the linearized Einstein equations
$\delta G_{AB}= 8\pi G \  \delta T_{AB}$
(such that $G_{AB}=\bar{G}_{AB} + \delta G_{AB}$ with
$\bar{G}_{AB}=0$), hold
\begin{eqnarray}
&& -3l^2 \left(p-1\right) ^2 \stackrel{\star\star}{\Psi} + 9l (p-p^2)
\stackrel{\star}{\Psi} +3t (3p+1) \dot\Psi +12 (p^2+p) \Psi -3t^2
\left( \frac{t_0}{t} \right) ^{2p} l^{\frac{-2}{p-1}} \nabla_r^2
\Psi\ \nonumber \\
&& = -16 \pi G \Psi  (p-1)^2 l^2 (\stackrel{\star}{\varphi_{(b)}})^2,
\label{ua} \\
&& l^2(p-1)^2 \stackrel{\star\star}{\Psi} + l(p^2-1)
\stackrel{\star}{\Psi} -3t^2 \ddot\Psi -5t (2p+1) \dot\Psi - 12p^2
\Psi +\frac{2}{3} t^2 \left( \frac{t_0}{t} \right) ^{2p}
l^{\frac{-2}{p-1}} \nabla_r^2 \Psi\ \nonumber \\
&& = 16 \pi G \Psi  t^2 (\dot\varphi_{(b)})^2,  \label{ub} \\
&& 3l (2p^2-3p+1) \stackrel{\star}{\Psi} -3t^2 \ddot\Psi - 15 p t
\dot\Psi + 12 p(1-2p) \Psi + t^2 \left( \frac{t_0}{t} \right) ^{2p}
l^{\frac{-2}{p-1}} \nabla_r^2 \Psi\ \nonumber \\
&& = 16 \pi G \Psi t^2 (\dot\varphi_{(b)})^2. \label{uc} 
\end{eqnarray}
Combinding the eqs. (\ref{ua}), (\ref{ub}) and (\ref{uc}),
we obtain an equation of motion for $\Psi$
\begin{eqnarray}
&& t^2\ddot\Psi -2(p-1)^2l^2 \stackrel{\star\star}{\Psi} +3(p+2)t
\dot\Psi +3(1-p)l\stackrel{\star}{\Psi} +12p\Psi - t^2 \left(
\frac{t_0}{t} \right) ^{2p} l^{\frac{-2}{p-1}} \nabla_r^2 \Psi\ \nonumber
\\
&& =-\frac{16}{3} \pi G \Psi \left[ t^2 (\dot\varphi_{(b)})^2 +l^2
(p-1)^2 (\stackrel{\star}{\varphi_{(b)}})^2 \right]. \label{aaa}
\end{eqnarray}

The Lagrange equation for $\Psi$:
${\partial ^{(5)} L \over
\partial \Psi } - {\partial \over \partial x^A} \left({\partial ^{(5)} L
\over \partial \Psi_{,A}}\right)=0$,
is given by
\begin{equation} \label{16}
\frac{\left(\dot\varphi\right)^2}{l^2} + \left(\frac{t}{t_0}\right)^{2p}
l^{\frac{2p}{p-1}} \left(\nabla_r\varphi\right)^2 +
\frac{(p-1)^2}{t^2} \left(\stackrel{\star}{\varphi}\right)^2 =0,
\end{equation}
where $^{(5)} L = \sqrt{\left|{^{(5)} g \over ^{(5)} g_0}\right|}
^{(5)} {\cal L}$.
In absence of the fluctuations $\delta \varphi$ and $\Psi$, the
eq. (\ref{16}) give us
\begin{equation}\label{2}
\frac{(\dot\varphi_{(b)})^2}{l^2}+\frac{(p-1)^2}{t^2}
(\stackrel{\star}{\varphi_{(b)}})^2=0.
\end{equation}
Furthermore, the Lagrange equation for $\varphi$:
${\partial ^{(5)} L \over
\partial \varphi}-{\partial \over \partial x^A} \left({\partial ^{(5)} L
\over \partial \varphi_{,A}}\right)=0$, is
\begin{eqnarray}
&&  (3p+1) \dot\varphi + t \ddot\varphi -
\frac{l^{\frac{-2}{p-1}} t_0^{2p}}{t^{2p-1}} \nabla_r^2\varphi -
(p-1)^2 \frac{l}{t} \left[ \frac{4p-1}{p-1}
\stackrel{\star}{\varphi} + l \stackrel{\star\star}{\varphi}
\right]  \nonumber \\
&&- 2\Psi \left\{ (3p+1) \dot\varphi + t \ddot\varphi +
\frac{l^{\frac{-2}{p-1}} t_0^{2p}}{t^{2p-1}} \nabla_r^2\varphi +
(p-1)^2 \frac{l}{t} \left[ \frac{4p-1}{p-1}
\stackrel{\star}{\varphi} + l \stackrel{\star\star}{\varphi}
\right] \right\} \nonumber \\
&&- 2 \left\{ t \dot\Psi \dot\varphi + \frac{l^{\frac{-2}{p-1}}
t_0^{2p}}{t^{2p-1}} \left[ \frac{\partial \Psi}{\partial x}
\frac{\partial\varphi}{\partial x} + \frac{\partial\Psi}{\partial
y} \frac{\partial\varphi}{\partial y} +
\frac{\partial\Psi}{\partial z} \frac{\partial\varphi}{\partial
z}\right] + (p-1)^2 \frac{l^2}{t} \stackrel{\star}{\Psi}
\stackrel{\star}{\varphi} \right\} =0. \label{19}
\end{eqnarray}
Notice that the equations (\ref{16}) and (\ref{19}) are very difficult
to be solved because $\varphi$ and $\Psi$ are quantum operators.
However, the eq. (\ref{16}) can be treated on the background
[see eq. (\ref{2})] and the eq. (\ref{19}) can be linearized using
the semiclassical expansion $\varphi(t,\vec r,l)=\varphi_{(b)}(t,l)
+ \delta\varphi(t,\vec r,l)$.
The linearized Lagrangian equations
take the form
\begin{eqnarray}
\ddot \varphi_{(b)} & + & \left( \frac{3p+1}{t}\right) \dot \varphi_{(b)} -
\frac{(p-1)^2 l^2}{t^2}\stackrel{\star\star}{\varphi_{(b)}} -
(p-1)(4p-1) \frac{l}{t^2} \stackrel{\star}{\varphi_{(b)}}=0, \label{back} \\
&&  \ddot \delta\varphi + \frac{3p+1}{t} \dot \delta\varphi
- \frac{(p-1)^2 l^2}{t^2}\stackrel{\star\star}{\delta\varphi} -
(p-1)(4p-1) \frac{l}{t^2} \stackrel{\star}{\delta\varphi} -
l^{\frac{-2}{p-1}} \left(\frac{t_0}{t}\right)^{2p}
\nabla_r^2\delta\varphi \nonumber \\
&& = 2\Psi\left[\ddot \varphi_{(b)} + \frac{3p+1}{t} \dot
\varphi_{(b)} + \frac{(p-1)^2
l^2}{t^2}\stackrel{\star\star}{\varphi_{(b)}} + (p-1)(4p-1)
\frac{l}{t^2} \stackrel{\star}{\varphi_{(b)}}
\right] \nonumber \\
&& + 2 \left[ \dot\Psi \dot\varphi_{(b)} + (p-1)^2 \frac{l^2}{t^2}
\stackrel{\star}{\Psi} \stackrel{\star}{\varphi_{(b)}} \right]. \label{3'}
\end{eqnarray}
The only solution of eq. (\ref{back}) on the manifold (\ref{3})
that is solution of
eq. (\ref{2}) is $\varphi_{(b)}(t,l)=C$, where $C$ is a constant. Hence,
the right side of eq. (\ref{3'}) becomes zero.
If we take into account the expression (\ref{2}), the eq. (\ref{aaa})
holds
\begin{equation} \label{21}
\ddot\Psi -2(p-1)^2\frac{l^2}{t^2} \stackrel{\star\star}{\Psi}
+\frac{3(p+2)}{t} \dot\Psi +3(1-p) \frac{l}{t^2}
\stackrel{\star}{\Psi} +12\frac{p}{t^2}\Psi - \left( \frac{t_0}{t}
\right) ^{2p} l^{\frac{-2}{p-1}} \nabla_r^2 \Psi =0.
\end{equation}
In order to simplify the structure of this equation we propose the
transformation $\Psi(t,\vec r,l) =
\left(\frac{t}{t_0}\right)^{-\frac{3(p+2)}{2}}
\left(\frac{l}{l_0}\right)^{-\frac{3}{4(p-1)}} \chi(t,\vec r,l)$,
such that the equation of motion for $\chi$ is
\begin{equation}
\ddot\chi -2(p-1)^2\frac{l^2}{t^2} \stackrel{\star\star}{\chi} -
\left( \frac{t_0}{t} \right) ^{2p} l^{\frac{-2}{p-1}} \nabla_r^2
\chi +\frac{1}{t^2} \left( -\frac{9}{4}p^2+3p+\frac{27}{8} \right)
\chi=0.
\end{equation}
We propose the following Fourier's expansion for $\chi$
\begin{equation}
\chi(t,\vec r,l) = \frac{1}{(2\pi)^{3/2}} {\Large\int} d^3 k_r
{\Large\int} dk_{l} \left[a_{k_r k_{l}} e^{i(\vec k_r.\vec r+
k_{l} l)} \xi_{k_rk_{l}}(t,l) + a^{\dagger}_{k_r k_{l}} e^{-i(\vec
k_r.\vec r+k_{l} l)} \xi^*_{k_r k_{l}}(t,l)\right],
\end{equation}
where the creation and annihilation operators
$a_{k_r k_{l}}$ and $a^{\dagger}_{k_r k_{l}}$
describe the algebra
\begin{displaymath}
\left[a_{k_rk_{l}}, a^{\dagger}_{k'_{r}k'_{l}}\right] =
\delta^{(3)}\left(\vec k_r - \vec k'_r \right) \delta\left( 
k_{l} - k'_{l}\right), \quad \left[a_{k_rk_{l}},
a_{k'_{r}k'_{l}}\right] = \left[a^{\dagger}_{k_rk_{l}},
a^{\dagger}_{k'_{r}k'_{l}}\right] =0.
\end{displaymath}

The dynamics of the modes $\xi_{k_rk_{l}}(t,l)$ is given
by
\begin{eqnarray}
&& \ddot{\xi}_{k_r k_{l}} + l^{-\frac{2}{p-1}}
\left(\frac{t_0}{t}\right)^{2p}
k^2_r \xi_{k_r k_l} + 2(p-1)^2 \frac{l^2}{t^2} \left\{
-\frac{\partial^2 }{\partial l^2} - 2 i k_l \frac{\partial}{\partial l}
\right. \nonumber \\
&& - \left.\frac{1}{2(p-1)^2 l^2}
\left(\frac{9}{4} p^2 - 3p -\frac{27}{8}\right) + k^2_l\right\} \xi_{k_r k_l}
=0.
\end{eqnarray}
Using the transformation
$\xi_{k_r k_l} = e^{-i \vec{k}_l.\vec{l}} \  \tilde\xi_{k_r k_l}$,
we obtain the following equation of motion for $\tilde\xi_{k_r k_l}$
\begin{equation}\label{4}
\ddot{\tilde\xi}_{k_r k_l} - 2(p-1)^2 \frac{l^2}{t^2}
\stackrel{\star\star}{\tilde\xi}_{k_r k_r}
+ \left[  l^{-\frac{2}{p-1}} \left(
\frac{t_0}{t}\right)^{2p} k^2_r - \left(\frac{9}{4} p^2 -3p-\frac{27}{8}
\right) \frac{1}{t^2} \right] \tilde\xi_{k_r k_l} =0.
\end{equation}
The problem with this equation is that it is not separable.
However, as we shall see later, this equation can be worked
in the limit $p \gg 1$, which is relevant for inflation.

\subsection{Inflationary case: $p \gg 1$}

In the limit $p \gg 1$ the equation (\ref{4}) can be simplified to
\begin{equation}
\ddot{\tilde\xi}_{k_r k_l} - 2 (p-1)^2 \frac{l^2}{t^2}
\stackrel{\star\star}{\tilde\xi}_{k_r k_r}
+\left[ k^2_r \left(\frac{t_0}{t}\right)^{2p} -
\left(\frac{9}{4} p^2 - 3p -\frac{27}{8}\right) \frac{1}{t^2} \right]
\tilde\xi_{k_r k_l} =0,
\end{equation}
which, once normalized, has the following solution:
\begin{equation}
\tilde\xi_{k_r k_l}(t,l) = A_1 \left[
B_1 \left(\frac{l}{l_0}\right)^{\frac{1+(1+4s^2)^{1/2}}{2}} +
B_2 \left(\frac{l}{l_0}\right)^{\frac{1-(1+4s^2)^{1/2}}{2}}\right]
\sqrt{\frac{t}{t_0}} {\cal H}^{(1)}_{\nu}[x(t)],
\end{equation}
where $\nu ={\sqrt{1+9p^2 -12 p -27/2+8(p-1)^2 s^2}\over 2(p-1)}$,
$x(t) = {k_r t^p_0\over t^{p-1} (p-1)}$ and $s^2$ is a separation
constant.

Furthermore, the only solution of (\ref{back}) that complies with
the expression (\ref{2}), is $\varphi_{(b)} = C$, where $C$ is a constant.
The equation of motion for the inflaton field (\ref{3'}),
for $ p \gg 1$, can be approximated to
\begin{equation} \label{3''}
\ddot{\delta\varphi} + \frac{(3p+1)}{t} \dot{\delta\varphi} -
(p-1)^2 \frac{l^2}{t^2} \stackrel{\star\star}{\delta\varphi} -
(p-1)(4p-1) \frac{l}{t^2} \stackrel{\star}{\delta\varphi} -
\left(\frac{t_0}{t}\right)^{2p} \nabla^2_r \delta\varphi =0.
\end{equation}
We can make the transformation
$\delta\varphi(t,\vec r,l) = \left({t\over t_0}\right)^{{-(3p+1)\over 2}}
\left({l\over l_0}\right)^{{-(4p-1)\over 2(p-1)}} \Pi(t,\vec r,l)$,
so that the equation of motion for $\Pi$ in the limit $p \gg 1$, is
\begin{equation}
\ddot\Pi - (p-1)^2 \frac{l^2}{t^2} \stackrel{\star\star}{\Pi} +
\frac{p}{2 t^2} \left(1-\frac{p}{2}\right) \Pi
-\left(\frac{t_0}{t}\right)^{2p} \nabla^2_r \Pi =0.
\end{equation}
The field $\Pi$ can be expressed as a Fourier expansion
in terms of its modes $\Pi_{k_r k_l}(t,\vec r,l) =
e^{i(\vec k_r . \vec r + k_l l)} \theta_{k_r k_l}(t,l)$,
such that the dynamics for $\theta_{k_r k_l}$ is described by
the equation
\begin{eqnarray}
\ddot{\theta}_{k_r k_l} &-& (p-1)^2 \frac{l^2}{t^2}
\stackrel{\star\star}{\theta}_{k_r k_l}
- 2i k_l (p-1)^2 \frac{l^2}{t^2}
\stackrel{\star}{\theta}_{k_r k_l} \nonumber \\
& + &
\left[ (p-1)^2 \frac{l^2}{t^2} k^2_l +
\frac{p}{2t^2} \left(1-\frac{p}{2}\right) + \left(\frac{t_0}{t}\right)^{2p}
k^2_r \right] \theta_{k_r k_l}=0,
\end{eqnarray}
which has the following normalized solution
\begin{equation}
\theta_{k_r k_l} = F_1 \  e^{-i k_l l}
\left[E_1
\left(\frac{l}{l_0}\right)^{\frac{1}{2}
+\frac{\sqrt{(p-1)^2 + 4q^2}}{2(p-1)}} + E_2
\left(\frac{l}{l_0}\right)^{\frac{1}{2}
-\frac{\sqrt{(p-1)^2 + 4q^2}}{2(p-1)}}\right] \sqrt{\frac{t}{t_0}}
{\cal H}^{(1)}_{\mu}[x(t)].
\end{equation}
Here, $\mu =
{\sqrt{(p-1)^2 + 4 q^2}\over 2(p-1)}={\sqrt{1+4 s^2}\over 2}$,
and ($F_1$, $E_1$, $E_2$,
$q^2=(p-1)^2 s^2$)
are constants.

\section{Effective 4D dynamics}

We consider the background metric (\ref{3}). If we take a foliation
such that $l = l_0$, the effective 4D background metric that results is
\begin{equation}\label{41}
dS^2_{(b)} \rightarrow ds^2_{(b)} = l^2_0 dt^2 -
\left(\frac{t}{t_0}\right)^{2p} l^{\frac{2p}{p-1}}_0 dr^2,
\end{equation}
and the effective 4D Lagrangian is $^{(4)} {\cal L}(\varphi,\varphi_{,\mu})
= {1\over 2} g^{\mu\nu} \varphi_{,\mu} \varphi_{,\nu} - V(\varphi)$,
such that the effective 4D background potential $V(\varphi_{(b)})$
induced on the metric (\ref{41}) is
\begin{equation}\label{pot}
V(\varphi_{(b)}) =\left. -
\frac{1}{2} \bar g^{ll} \left[\varphi_{(b),l}\right]^2\right|_{l=l_0}
=\left.\frac{1}{2} \frac{(p-1)^2}{t^2}
\left[\varphi_{(b),l}\right]^2 \right|_{l=l_0}.
\end{equation}
In the limit $p \gg 1$, which is relevant for inflation,
the metric (\ref{41}) can be approximated to
\begin{equation}\label{42}
ds^2_{(b)} = l^2_0 dt^2 - \left(\frac{t}{t_0}\right)^{2p} l^2_0 dr^2.
\end{equation}
The effective 4D potential (\ref{pot}) can be founded by solving
the effective equation of motion for $\varphi_{(b)}$
\begin{equation}\label{mo}
\left.\ddot\varphi_{(b)} + \frac{(3p+1)}{t} \dot\varphi_{(b)}
- \frac{(p-1)^2}{t^2} l^2 \stackrel{\star\star}{\varphi_{(b)}}
- (p-1)(4p-1) \frac{l}{t^2} \stackrel{\star}{\varphi_{(b)}}\right|_{l=l_0}=0,
\end{equation}
on the effective background metric (\ref{42}). If we make
$\varphi_{(b)}(t,l) = \varphi_1(t) \varphi_2(l)$, we obtain
\begin{eqnarray}
&& \ddot\varphi_1 + \frac{(3p+1)}{t} \dot\varphi_1 = -\frac{m^2}{t^2}
\varphi_1, \\
&& l^2 (p-1)^2 \stackrel{\star\star}{\varphi_2} +
(p-1) (4p-1) l \stackrel{\star}{\varphi_2} = -m^2 \varphi_2.
\end{eqnarray}
The general solutions for these equations are
\begin{eqnarray}
\varphi_1(t) &=& t^{-3p/2} \left[ B_1 t^{\frac{\sqrt{9p^2 - 4m^2}}{2}}
+ B_2 t^{-\frac{\sqrt{9p^2-4m^2}}{2}}\right], \\
\varphi_2(l) & = &
l^{\frac{-3p}{2(p-1)}} \left[ A_1 l^{\frac{\sqrt{9p^2 -4m^2}}{2(p-1)}}
 + A_2 l^{-\frac{\sqrt{9p^2 -4m^2}}{2(p-1)}}\right],
\end{eqnarray}
where $A_1$, $A_2$, $B_1$, $B_2$ are constants of integration and $m$
is a separation constant. If we choose $A_1=B_2=m=0$, we obtain
that $\varphi_1(t) = B_1$ and $\varphi_2(l)=A_2 l^{3p/(p-1)}$, such that
${\partial\varphi_{(b)}\over\partial t}=0$ and ${\partial\varphi_{(b)}\over
\partial l}= {-3p \over (p-1)} l^{-1} \varphi_{(b)}$.
Therefore, the induced 4D background potential will be
\begin{equation}\label{bpot}
\left.V(\varphi_{(b)})\right|_{l=l_0} = \left.
\frac{9p^2}{2 t^2 l^2} \varphi^2_{(b)} \right|_{l=l_0} =
\frac{9p^2}{2 t^2 l^2_0} \varphi^2_{(b)}.
\end{equation}
The metric (\ref{42}), with the change of variables
$\tau=l_0 t$ and $R=l_0 r$,
becomes
\begin{equation}\label{43}
ds^2_{(b)} = d\tau^2 - \left(\frac{\tau}{\tau_0}\right)^{2p} dR^2.
\end{equation}
Due to the fact ${\partial \varphi_{(b)}\over \partial\tau} =0$,
the effective 4D background energy density is
\begin{equation}
\rho_b = \left.\frac{1}{2} \left(\frac{\partial\varphi_{(b)}}{\partial\tau}
\right)^2 + V(\varphi_{(b)})\right|_{l=l_0}
 =\left. \frac{9 p^2}{2\tau^2} \varphi^2_{(b)}\right|_{l=l_0}
=\frac{3 H^2(\tau)}{8\pi G}.
\end{equation}
Here, the Hubble parameter $H(\tau) = {1 \over a} {da\over d\tau} =p/\tau$,
which is related
to the scale factor $a(\tau) = a_0 \left(\tau/\tau_0\right)^{2p}$
and $\varphi_{(b)}(\tau,l=l_0)$ is a constant: $\varphi_{(b)}=
{1\over 2\sqrt{3\pi G}}$.

\subsection{Effective 4D dynamics of metric fluctuations}

Using the fact that the solution of eq. (\ref{21}), in the
limital case $p \gg 1$, can be written as
\begin{equation}
\Psi(t, \vec r, l) = \left[ B_1
\left(\frac{l}{l_0}\right)^{\frac{2p-5}{4(p-1)}+\frac{\sqrt{1+4 s^2}}{2}}
+ B_2 \left(\frac{l}{l_0}\right)^{\frac{2p-5}{4(p-1)}
-\frac{\sqrt{1+4 s^2}}{2}}
\right] \psi(t,\vec r),
\end{equation}
and that on the hypersurface $l=l_0$ one obtains
\begin{eqnarray}
&& \Psi(t,\vec r, l=l_0) = (B_1 + B_2) \psi(t,\vec r), \\
&& \left.\frac{\partial\Psi}{\partial l}\right|_{l=l_0} =
\left[\frac{2p-5}{4(p-1)} \frac{(B_1 + B_2)}{l_0} +
\frac{\sqrt{1+4 s^2}}{2} \frac{(B_1-B_2)}{l_0} \right]
\psi(t,\vec r), \\
&& \left.\frac{\partial^2 \Psi}{\partial l^2}\right|_{l=l_0} =
\left[ \left(\frac{9}{16(p-1)^2} + s^2\right) \frac{(B_1 + B_2)}{l^2_0}
-\frac{3}{4} \frac{\sqrt{1+4 s^2}}{(p-1)} \frac{(B_1 - B_2)}{l^2_0}
\right] \psi(t, \vec r).
\end{eqnarray}
Hence,
the effective equation of motion for $\psi(\tau=l_0 t,\vec R = l_0 \vec r)$
on the hypersurface $l=l_0$ (which only is valid for $p\gg 1$),
will be
\begin{equation}\label{44}
\frac{\partial^2 \psi}{\partial\tau^2} +
\frac{3(p+2)}{\tau}\frac{\partial \psi}{\partial\tau}
+ \frac{\psi}{\tau^2} \left[
\frac{21}{8} + \frac{21 p}{2} - 2(p-1)^2 s^2 + \frac{4 p^2}{l^2_0}
\right] - \left(\frac{\tau_0}{\tau}\right)^{2p} \nabla^2_R \psi =0,
\end{equation}
when we have used $B_1=B_2$.
The equation (\ref{44}) can be simplified using the transformation
$\psi(\tau,\vec R,l_0) = \left(\frac{\tau}{\tau_0}\right)^{\frac{
-3(p+2)}{2}} \chi(\tau,\vec R)$, such that the equation of motion
for $\chi$ is
\begin{equation}\label{45}
\frac{\partial^2 \chi}{\partial\tau^2} + \frac{\chi}{\tau^2} \left[
\frac{27}{2} p + \frac{69}{8} - 2(p-1)^2 s^2 + \frac{4p^2}{l^2_0}\right]
-\left(\frac{\tau_0}{\tau}\right)^{2p} \nabla^2_R \chi =0.
\end{equation}
The field $\chi(\tau,\vec R)$ can be expanded in terms of their
modes $\chi_{k_R}(\tau, \vec R) =
e^{i \vec{k_R}.\vec R} \tilde\xi_{k_R}(\tau)$, and the equation of motion
for the $\tau$-dependent modes $\tilde\xi_{k_R}(\tau)$, is
\begin{equation}\label{46}
\frac{\partial^2 \tilde\xi_{k_R}}{\partial\tau^2} +
\left[k^2_R \left(\frac{\tau_0}{\tau}\right)^{2p} -
\frac{\left[2(p-1)^2 s^2 - \frac{4p^2}{l^2_0} -
\left(\frac{27 p}{2} + \frac{69}{8}\right)\right]}{\tau^2} \right]
\tilde\xi_{k_R}=0.
\end{equation}
The normalized solution for this equation is
\begin{equation}                        \label{47}
\tilde\xi_{k_R} (\tau, \vec R) = A \sqrt{\frac{\tau}{\tau_0}}
{\cal H}^{(1)}_{\nu}
[x(\tau)],
\end{equation}
where ${4(p-1)\over \tau_0 \pi} |A|^2=1$,
${\cal H}_{\nu}[x(\tau)]$ is the first kind Hankel function
with argument $x(\tau) = k_R {\tau^p_0 \over (p-1) \tau^{p-1}}$ and
$\nu = {\sqrt{8(p-1)^2 s^2 - {16 p^2\over l^2_0} -54 p -{67\over 2}}\over
2(p-1)}$.

\subsection{Energy density fluctuations}

The energy density fluctuations on the effective 4D FRW metric is\cite{ant}
\begin{equation}
\left.\frac{\delta\rho}{\left<\rho\right>}\right|_{IR} \simeq 2 \Psi,
\end{equation}
where the brackets $<...>$ denote the expectation value on the
3D hypersurface $\vec R(X,Y,Z)$. This approximation is valid during
inflation on super Hubble scales. The amplitude for the 4D gauge-invariant
metric fluctuations on cosmological scales is
\begin{equation}
\left.\left<\Psi^2\right> \right|_{IR} \simeq 4 B^2_1 \left<\psi^2\right>
= 4 B^2_1 \left(\frac{\tau}{\tau_0}\right)^{-3(p+2)} \left.\left<
\chi^2\right>\right|_{IR},
\end{equation}
so that
\begin{equation}
\left.\left<\Psi^2\right>\right|_{IR} \simeq
\frac{4 B^2_1\left(\frac{\tau}{\tau_0}\right)^{-3(p+2)}
}{(2\pi)^3} \int^{\epsilon (p/\tau)(\tau/\tau_0)^p}_0
d^3 k_R \tilde\xi_{k_R} \tilde\xi^*_{k_R} =
\frac{1}{(2\pi)^3} \int^{\epsilon (p/\tau)(\tau/\tau_0)^p}_0
\frac{dk_R}{k_R} {\cal P}_{\Psi}(k_R),
\end{equation}
being $\epsilon \simeq 10^{-3}$ a dimensionless constant and
${\cal P}_{\Psi}(k_R) \sim k^{3-2\nu}_R$
the power spectrum of $\left<\Psi^2\right>$.
It is known from experimental data\cite{PDG}, that the spectral $n_s$
index for this spectrum is
\begin{equation}         \label{condi}
n_s = 0.97 \pm 0.03,
\end{equation}
where, in our case, $n_s = \left. 4-\sqrt{8(p-1)^2 s^2 - {16 p^2\over l^2_0}
-54 p - {67 \over 2}}/(p-1)\right|_{p\gg 1} \simeq
4- \sqrt{8(s^2 - 2/l^2_0)}$.
Hence, from the condition (\ref{condi}), we obtain
\begin{equation}\label{lala}
\frac{9}{8} + \frac{2}{l^2_0} < s^2 <  \frac{(3.06)^2}{8} + \frac{2}{l^2_0},
\end{equation}
which provide us a cut for the separation constant $s$ in terms of the
value of the fifth coordinate $l_0$. Note that for each possible
foliation $l^{(i)}_0$ there is a range of possible separation constants
$s^{(i)}$.

\subsection{Effective 4D dynamics of the inflaton field
fluctuations}

If we take into account that $\tau = l_0 t$ and $R=l_0 r$, the
5D equation (\ref{3'}) for $p \gg 1$, holds [see eq. (\ref{3''})]
\begin{equation}
\frac{\partial^2\delta\varphi}{\partial\tau^2} + \frac{(3p+1)}{\tau}
\frac{\partial\delta\varphi}{\partial\tau} -
(p-1)^2 \frac{l^2}{\tau^2} \frac{\partial^2 \delta\varphi}{\partial l^2}
-(p-1)(4p-1) \frac{l}{\tau^2} \frac{\partial\delta\varphi}{\partial l} -
\left(\frac{\tau_0}{\tau}\right)^{2p} \nabla^2_R \delta\varphi =0.
\end{equation}
Note that for this limital case $\delta\varphi$ is independent of $\Psi$.
Using the fact that
\begin{eqnarray}
&& \left. \delta\varphi\right|_{l=l_0} =
(E_1 + E_2) \  {\cal G}(\tau,\vec R), \\
&& \left. \frac{\partial \delta\varphi}{\partial l}\right|_{l=l_0} =
\frac{1}{2 l_0 (p-1)} \left[ -3 p (E_1 + E_2) +
\sqrt{(p-1)^2 + 4 q^2} (E_1 - E_2)\right] {\cal G}(\tau,\vec R), \\
&& \left. \frac{\partial^2 \delta\varphi}{\partial l^2}\right|_{l=l_0} =
\frac{1}{4 l_0^2 (p-1)^2} \left\{\left[(4p-1)^2 + 4q^2\right]
(E_1+ E_2) \right. \nonumber \\
&+& \left. (2-8p) \sqrt{(p-1)^2 + 4 q^2}
(E_1 - E_2) \right\} {\cal G}(\tau, \vec R),
\end{eqnarray}
we obtain, for $E_1=E_2$,
the equation of motion for the function ${\cal G}$
\begin{equation}
\frac{\partial^2 {\cal G}}{\partial\tau^2} + \frac{(3p+1)}{\tau}
\frac{\partial {\cal G}}{\partial \tau} +
+ \frac{{\cal G}}{\tau^2} \left[ 2p^2 + \frac{p}{2} -\left(
\frac{1}{4} + q^2\right) \right] - \left(\frac{t_0}{t}\right)^{2p}
\nabla^2_R {\cal G} =0.
\end{equation}
Now we can make the following transformation: ${\cal G}(\tau,\vec R) =
\left({\tau\over \tau_0}\right)^{{-(3p+1)\over 2}} \vartheta(\tau,\vec R)$.
Hence, the equation of motion for the $\tau$-dependent
modes $\tilde\vartheta_{k_R}(\tau)$ of $\vartheta(\tau,\vec R)$, holds
\begin{equation}
\frac{\partial^2\tilde\vartheta_{k_R}}{\partial\tau^2}+
\left[\left(\frac{\tau_0}{\tau}\right)^{2p} k^2_R -
\frac{\left(\frac{p^2}{4} + q^2 -\frac{p}{2}\right)}{\tau^2}\right]
\tilde\vartheta_{k_R} =0.
\end{equation}
The normalized solution for the time dependent modes
$\vartheta_{k_R}(\tau)$ is
\begin{equation}
\vartheta_{k_R}(\tau) =
K_1 \sqrt{\frac{\tau}{\tau_0}} \  {\cal H}^{(1)}_{\sigma}[x(\tau)],
\end{equation}
where $\sigma = {\sqrt{(p-1)^2 + 4 q^2} \over 2(p-1)}
={\sqrt{1+4 s^2}\over 2}$,
$x(\tau) = {k_R \tau^p_0 \over (p-1) \tau^{p-1}}$ and
${4 (p-1) \over \tau_0 \pi} |K_1|^2 =1$.

The squared $\delta\varphi$-fluctuations are
\begin{equation}
\left.\left<\delta\varphi^2\right>\right|_{IR} \simeq
\frac{4 E^2_1}{(2\pi)^3} \left(\frac{\tau}{\tau_0}\right)^{-(3p+1)}
\int^{\epsilon (p/\tau) (\tau/\tau_0)^p}_0 d^3k_R \vartheta_{k_R}(\tau)
\vartheta^*_{k_R}(\tau) = \frac{1}{(2\pi)^3}
\int^{\epsilon (p/\tau) (\tau/\tau_0)^p}_0 {\cal P}_{\delta\varphi}(k_R),
\end{equation}
where ${\cal P}_{\delta\varphi} \sim k^{3-2\sigma}_{R}$ is the power-spectrum
of $\left.\left<\delta\varphi^2\right>\right|_{IR}$.
Taking into account the condition (\ref{lala}), we obtain
\begin{equation}
\frac{\sqrt{1+\frac{9}{2}+\frac{8}{l^2_0}}}{2} < \sigma <
\frac{\sqrt{1+\frac{(3.06)^2}{2}+\frac{8}{l^2_0}}}{2},
\end{equation}
such that, for a scale invariant power-spectrum ${\cal P}_{\delta\varphi}$
(i.e., for $\sigma=3/2$), we obtain
\begin{equation}
2.2857 < l^2_0 < 2.4109,
\end{equation}
which is a good estimation of $l_0$ for a nearly scale invariant
${\cal P}_{\delta\varphi}$ obtained from the experimental data (\ref{condi}).\\

\section{Final Comments}

In this letter we have studied 4D gauge-invariant
(scalar) metric fluctuations in space time matter inflation,
using the 5D Ponce de Leon background metric. This metric describes
a 5D universe in an apparent vacuum $\bar{G}_{AB}=0$, but however
it is not Riemann flat.
In general (i.e., for
an arbitrary power of expansion $p$), the equations of motion for
$\delta\varphi$ and $\Psi$ are not factorizable, so that
its treatment is very difficult. However, in the limital case
$p \gg 1$ the treatment is possible because both, $\delta\varphi(t,\vec r,l)$
and $\Psi(t, \vec r, l)$ can be written as product of functions
of $t$, $\vec r$ and $l$ (i.e., both functions are factorizable). In such
a case we have found that $\Psi$ and $\delta\varphi$ become
independents (as in an effective 4D de Sitter expansion studied
in \cite{ant}), because the effective 4D background field
$\varphi_{(b)}$ becomes a constant of $\tau=l_0 t$. 
This is not surprasing because the case $p \gg 1$ describes
an effective 4D
asymptotic de Sitter expansion. In particular, using experimental
values of $n_s$ (for the
$\delta\rho/\rho$-spectrum), we have found that
the fifth coordinate used for the foliation $l=l_0$
could take values close to $l_0 \simeq 1.55$
in a scale invariant ${\cal P}_{\delta\varphi}$-power spectrum.\\

\noindent
{\bf Acknowledgments}\\
M.A. and M. B. acknowledge Universidad Nacional de Mar del Plata and
CONICET (Argentina) for financial support.\\

\end{document}